\documentclass[aps,pre,floatfix,twocolumn,a4paper,superscriptaddress]{revtex4}

\usepackage{amsmath}
\usepackage{afterpage}
\usepackage{amsfonts}
\usepackage{graphicx}
\usepackage{epsfig}
\usepackage{color}
\usepackage{xspace}
\usepackage{subfigure}

\newcommand{\ket}[1]{ \ensuremath{\left| #1 \right\rangle} }

\newcommand{\EX}[1] { \ensuremath{\left\langle #1 \right\rangle} }
\newcommand{\half} {\ensuremath{\frac{1}{2}}}
\newcommand{\ihbar}{\ensuremath{\frac{i}{\hbar}}}

\newcommand{\BUE}{The British University in Egypt,
                  El Sherouk City, 
                  Misr  Ismalia Desert Road,
                  Postal No. 11837
                  P.O. Box 43,
                  Egypt.
}
\newcommand{\Sussex}{Centre for Physical Electronics and Quantum Technology, 
                      School of Science and Technology, 
                      University of Sussex, 
                      Falmer, 
                      Brighton, 
                      BN1 9QT, 
                      UK.}
\newcommand{\SPAWAR}{Space and Naval Warfare Systems Center, 
                      Code 2363, 
                      53560 Hull Street,
                      San Diego, 
                      California 92152-5001, 
                      USA.}
\newcommand{\Liverpool}{Department of Electrical and Electronic Engineering, 
                         Liverpool University,
                         Brownlow Hill, 
                         Liverpool, 
                         L69 3GJ, 
                         UK.}

\begin{document}

\title{Signatures of chaotic and non-chaotic-like behaviour in a
non-linear quantum oscillator through photon detection.}
\author{M.J.~Everitt}
\email{m.j.everitt@physics.org}
\affiliation{\BUE}
\affiliation{\Sussex}
\author{T.D.~Clark}
\affiliation{\Sussex}
\author{P.B.~Stiffell}
\affiliation{\Sussex}
\author{J.F.~Ralph}
\affiliation{\Liverpool}
\author{A.R.~Bulsara}
\affiliation{\SPAWAR}
\author{C.J.~Harland}
\affiliation{\Sussex}

\begin{abstract}
  The  driven non-linear duffing  oscillator is a  very good,
  and  standard, example  of a  quantum mechanical  system  from which
  classical-like  orbits can  be  recovered from  unravellings of  the
  master  equation.  In order  to generate  such trajectories  in the
  phase space  of this oscillator in  this paper we use  a the quantum
  jumps  unravelling  together  with  a suitable  application  of  the
  correspondence  principle.   We  analyse  the  measured  readout  by
  considering  the power  spectra  of photon  counts  produced by  the
  quantum jumps.   Here we show  that localisation of the  wave packet
  from  the  measurement of  the  oscillator  by  the photon  detector
  produces  a  concomitant  structure  in  the power  spectra  of  the
  measured  output.  Furthermore,  we demonstrate  that  this spectral
  analysis can be  used to distinguish between different  modes of the
  underlying dynamics of the oscillator.
\end{abstract}

\maketitle

\section{Introduction}
There is  currently an  intense interest being  shown in  the possible
application  of  quantum  devices  to  fields such  as  computing  and
information   processing~\cite{Nie00}.  The   goal  is   to  construct
machinery  which operates  manifestly  at the  quantum  level. In  any
successful development  of such technology the role  of measurement in
quantum  systems will be  of central,  indeed crucial,  importance (see for example~\cite{shepelyansky01}). In
order to  extend our  understanding of this  problem we  have recently
investigated the coupling together of  quantum systems that, to a good
approximation,  appear classical  (via the  correspondence  limit) but
whose      underlying      behaviour      is     strictly      quantum
mechanical~\cite{everitt05}. In this work we followed the evolution of
two  coupled,  and identical,  quantised  Duffing  oscillators as  our
example system.   
We utilised two unravellings of  the master equation
to  describe this system:  quantum state  diffusion and  quantum jumps
which   correspond,   respectively,   to  unit-efficiency   heterodyne
measurement  (or   ambi-quadrature  homodyne  detection)   and  photon
detection~\cite{Wis96}.  
We demonstrated  that  the entanglement  that
exists  between the  two oscillators  depends on  the nature  of their
dynamics.  Explicitly,   
we  showed  that  whilst   the  dynamics  was
chaotic-like the  entanglement between the  oscillators remained high;
conversely, if the two oscillators entrained into a periodic orbit the
degree of entanglement became very small.

With this background we subsequently became interested in acquiring a
detailed understanding of experimental readouts of quantum
chaotic-like systems. In this paper we have chosen to explore the
subject through the quantum jumps unravelling of the master
equation~\cite{Ple98,Wis96,Heg93}.  Here, the measured output is
easily identified, namely a click or no click in the photon detector.
However, this measurement process is rather unique in the fact that it
possesses no classical analogue. Indeed, this is the case even when
the system under consideration may appear to be evolving along a
classical trajectory. Interestingly, despite the fact that the photon
detector has no classical analogue, it is the very presence of this as
a source of decoherence that is responsible for recovering
classical-like orbits in the $\left( \left\langle q\right\rangle
,\left\langle p\right\rangle \right) $ phase plane (despite the fact
that we measure neither $q$ nor $p$).  The subject of recovering such
chaotic-like dynamics from unravellings of the master equation has
been studied in depth in the
literature~\cite{Per98,Hab98,brun97,Bru96,Spi94} and a detailed
discussion is beyond the scope of this paper. However, we note that
recently in~\cite{peano04} resonances have been observed in a model of
a non-linear nano-mechanical resonator that is absent in the
corresponding classical model. In this present work we have chosen to
scale the oscillator so that we recover orbits similar to those
generated from a classical analysis.

\section{Background}

In this work we study the output resulting from the measurement of
quantum objects where the measurement device generates decoherence effects.
In this limit the system exhibits dynamical behaviour in terms of its
expectation values very much like those observed in its classical
counterpart.  In this work we investigate the region of parameter
space under which the classical system exhibits chaotic motion.  Of
the many models that could be used we have chosen the Quantum Jumps
approach~\cite{Ple98,Wis96,Heg93}.  We note that this is only one of
several possible unravellings of the master equation that correspond
to the continuous measurement of the quantum object considered. Our
motivation for using this approach is that the recorded output of the
measurement is completely transparent i.e.  the photon counter either
registers a photon or it does not.

In the quantum jumps unravelling  of the master equation the evolution
of the  (pure) state  vector $\left\vert \psi  \right\rangle $  for an
open  quantum system  is  given by  the  stochastic It\^{o}  increment
equation
\begin{widetext}
  \begin{equation}
    \label{eq:jumps}
    \ket{d \psi}
     =  - \ihbar H   \ket{\psi} dt 
        - \half \sum_j \left[L_j^\dag L_j - \EX{L_j^\dag L_j} \right] \ket{\psi} dt 
        + \sum_j \left[ \frac{L_j}{\sqrt{\EX{L_j^\dag L_j}}} - 1 \right] \ket{\psi} dN_j
  \end{equation}
\end{widetext}
where $H$ is  the Hamiltonian, $L_{i}$ are the  Linblad operators that
represent coupling  to the environmental  degrees of freedom,  $dt$ is
the time  increment, and $dN_{j}$  is a Poissonian noise  process such
that     $dN_{j}dN_{k}=\delta    _{jk}dN_{j}$,     $dN_{j}dt=0$    and
$\overline{dN_{j}}=\left\langle  L_{j}^{\dag }L_{j}\right\rangle  dt$. 
These latter conditions imply that jumps occur randomly at a rate that
is determined  by $\left\langle L_{j}^{\dag  }L_{j}\right\rangle $. We
will  find that  this is  very important  when explaining  the results
presented later in this paper.
For an excellent discussion  of quantum trajectories interpreted as a
a realistic  model of  a system that  is being  continuously monitored
see~\cite{Wis96}. For  an interesting and more  general discussion on
the emergence of classical-like behaviour from quantum systems
see~\cite{Giulini96,joos03}.

The Hamiltonian  for our, standard, example  system of the Duffing  oscillator is
given by
\begin{equation}
H = \frac{1}{2} p^2 + \frac{\beta^{2}}{4}q^4 - \frac{1}{2} q^2 + \frac{g}{\beta}\cos\left( t \right) q+\frac{\Gamma}{2}\left( q p + p q \right)
\label{ham}
\end{equation}
where $q$ and $p$ are  the canonically conjugate position and momentum
operators for the oscillator. In this example we have only one Linblad
operator which  is $L=\sqrt{2\Gamma} a$,  where $a$ is  the oscillator
annihilation  (lowering)  operator, $g$  is  the  drive amplitude  and
$\Gamma=0.125$ quantifies the damping.

In order  to apply  the correspondence principal  to this  system, and
recover    classical-like   dynamics,    we    have   introduced    in
Eq.~(\ref{ham}) the  parameter $\beta $. For  this Hamiltonian it
has two  interpretations that are  mathematically equivalent. Firstly,
it can  be considered to scale  $\hbar $ itself,  or, alternatively we
can simply view $\beta $  as scaling the Hamiltonian, leaving $\hbar $
fixed, so  that the relative motion  of the expectation  values of the
observables becomes large compared with the minimum area $\left( \hbar
  /2\right) $ in  the phase space. In either  case, the system behaves
more classically as  $\beta $ tends to zero from  its maximum value of
one. In this work we have chosen to set $\beta =0.1$.

\section{Results}
 \begin{figure}[tbh]
 \begin{center}
 \subfigure[Classical Duffing oscillator.\label{fig:powerCX}]{
   \resizebox*{0.45\textwidth}{!}{\includegraphics{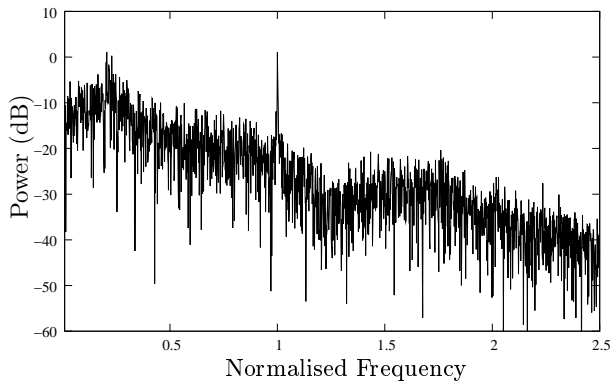}}
 }
 \subfigure[Quantum Duffing oscillator.\label{fig:powerQX}] {
   \resizebox*{0.45\textwidth}{!}{\includegraphics{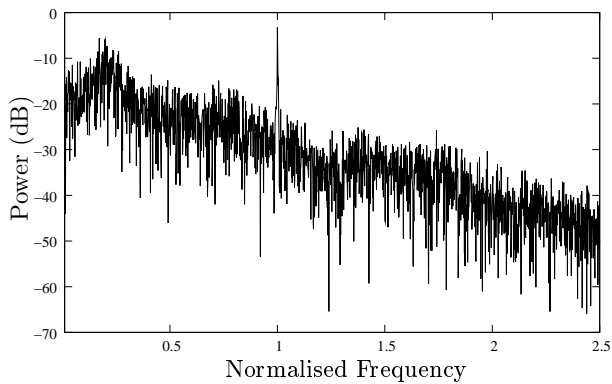}}
 }
 \caption{
   Power  spectrum of  the  position, $x$  for  the classical  Duffing
   oscillator  and   $\EX{q}$  for  the   quantum  Duffing  oscillator
   $\beta=0.1$. The frequency is  normalised to the drive frequency of
   the oscillator.
 \label{fig:powerCQX}}
 \end{center}
\end{figure}
Let us now consider the  specific example of a Duffing oscillator with
a drive amplitude $g=0.3$. This parameter, together with all the those
already specified,  form the classic example used  to demonstrate that
chaotic-like behaviour  can be recovered  for open quantum  systems by
using           unravellings          of           the          master
equation~\cite{Per98,brun97,Bru96,everitt05}.                        In
Fig.~\ref{fig:powerCQX} we compare the  power spectra of the classical
position coordinate with that  of $\left\langle q\right\rangle $. Here
noise has been added to the  classical system so as to mimic the level
of quantum  noise that  is present in  the stochastic elements  of our
chosen unravelling  of the  master equation and  we have solved  for a
realisation of the  Lagivan equation.  As can be  seen, for this value
of $\beta  $ there  is a very  good match  between these two  results. 
Moreover, both display power  spectra that are typical for oscillators
in chaotic orbits.

However, it is not position that is the measured output in this model,
but the quantum jumps recorded, as a function $\mathcal{N}(t)$ of time
in the photon  detector.  As stated above, these  jumps occur randomly
at   a  rate   that   is  determined   by  $\left\langle   L_{j}^{\dag
  }L_{j}\right\rangle   $  which,  for   this  example,   is  $2\Gamma
\left\langle  n\right\rangle $.   Hence, the  probability of  making a
jump is  proportional to  the number  of photons in  the state  of the
system at  any one time.  

\begin{figure}[tbh]
 \begin{center}
 \subfigure[Power spectrum of $\EX{q}$.\label{fig:powerShoX}]{
   \resizebox*{0.45\textwidth}{!}{\includegraphics{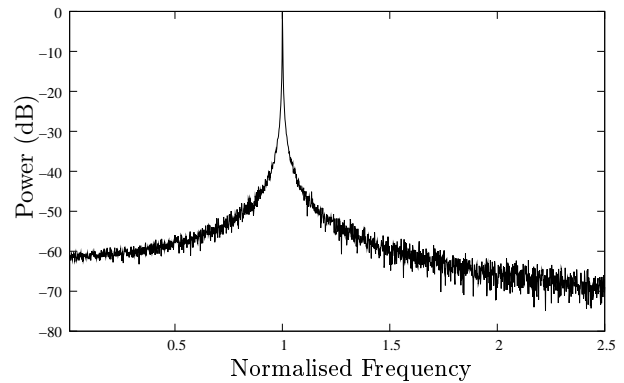}}
 }
 \subfigure[Power spectrum of $\mathcal{N}(t)$.\label{fig:powerShoN}] {
   \resizebox*{0.45\textwidth}{!}{\includegraphics{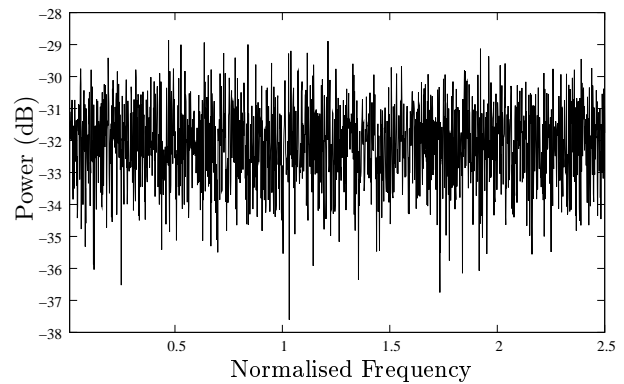}}
 }
 \caption{
   Power spectrum of the position $\EX{q}$ and photons counted $\mathcal{N}(t)$ for the quantum
   simple harmonic oscillator in a steady state. Here the frequency is
   normalised to the drive frequency of the oscillator.
 \label{fig:powerSho}}
 \end{center}
\end{figure}
We now consider a special case that occurs frequently in the classical
limit,  namely where  $\left\vert  \psi \right\rangle  $ localises  
approximately to a coherent (Gaussian) state.  It is apparent that for such
a state the  chance of observing a jump is  proportional to the square
of the  distance in $\left(  \left\langle q\right\rangle ,\left\langle
    p\right\rangle \right) $  of the state from the  origin.  In order
to  illustrate the  implications of  this,  let us  consider a  driven
simple harmonic oscillator. The Hamiltonian is
$$
H_s = \frac{1}{2} p^2 + \frac{1}{2} q^2 + \frac{g}{\beta}\cos\left( t \right) q
$$
and we note that in  this special case the only effect of $\beta=0.1$ is
 to  scale the  amplitude of the drive (again  we set $g=0.3$).  We now
solve Eq.~(\ref{eq:jumps}) using this Hamiltonian and allow the system
to settle into  a steady state.  Then, as the  phase portrait for this
system simply describes a circle centred about $(0,0)$ we would expect
the power spectra of photons counted to be the same as those for white
noise. Indeed,  this is clearly seen  in Fig.~\ref{fig:powerSho} where
we show the power spectrum for  both (a) the position operator and (b)
the measured quantum jumps.

\begin{figure}[tbh]
 \begin{center}
 \resizebox*{0.45\textwidth}{!}{\includegraphics{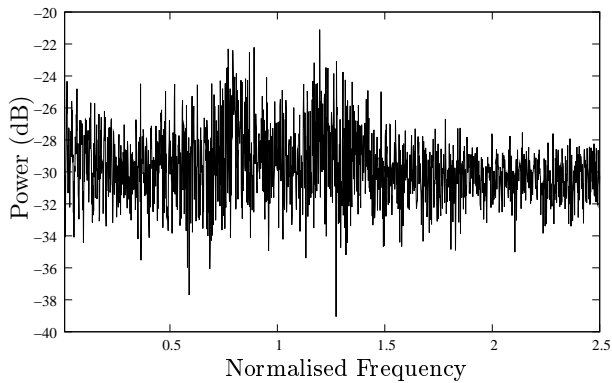}}
 \caption{
   Power  spectrum  of the  measured  quantum  jumps $\mathcal{N}(t)$ for the  Duffing
   oscillator of Fig.~\ref{fig:powerQX}.
 \label{fig:powerQJ}}
 \end{center}
\end{figure}

For more  complicated orbits, such  as those exhibited by  the Duffing
oscillator, we  would expect  to see some  evidence of  the underlying
dynamical   behaviour.   Hence,   localisation  of   $\left\vert  \psi
\right\rangle $ from the measurement of the Duffing oscillator through
the  photon  detector  forms  a  concomitant structure  in  the  power
spectrum of the measured output. In Fig.~\ref{fig:powerQJ}  we show, for comparison with~\ref{fig:powerQX}, such a power spectrum.

\begin{figure}[tbh]
 \begin{center}
 \resizebox*{0.45\textwidth}{!}{\includegraphics{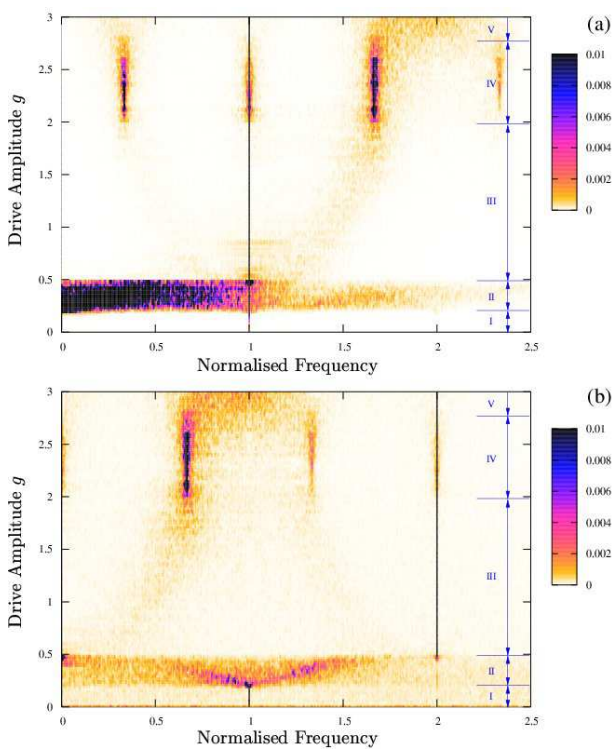}}
 \caption{
(color online)   Power   spectrum  of  the   \textbf{(a)}  $\EX{q}$  and
   \textbf{(b)}  measured  quantum  jumps   as  a  function  of  drive
   amplitude.
 \label{fig:power3d}}
 \end{center}
 \end{figure}

 As we  can see from  Fig.~\ref{fig:powerQJ} the power  spectrum for
 this chaotic mode of operation reveals some structure. However, it is
 not clear from this picture alone  how we might relate this result to
 that shown in Fig.~\ref{fig:powerQX}. It is therefore reasonable to
 ask if this result does  indeed tell us anything about the underlying
 dynamics of the oscillator. We have addressed this point by computing
 the  power  spectrum  of  both  $\left\langle  q\right\rangle  $  and
 $\mathcal{N}(t)$ for  drive amplitudes in the range  $0<g\leq 3$, the
 results of which are presented in Fig.~\ref{fig:power3d}.

 Although the functional form of these power spectra obviously differ,
 they do clearly  exhibit changes in behaviour that  are coincident in
 the  drive  amplitudes  of  both  figures. These  are  identified  as
 intervals in $g$ labelled I,~II,~\ldots in Fig.~\ref{fig:power3d}.

 \begin{figure*}[!p]
 \begin{center}
 \subfigure[Power spectrum of $\EX{q}$, $g=0.1$.\label{fig:ps0.1x}]{
   \resizebox*{0.45\textwidth}{!}{\includegraphics{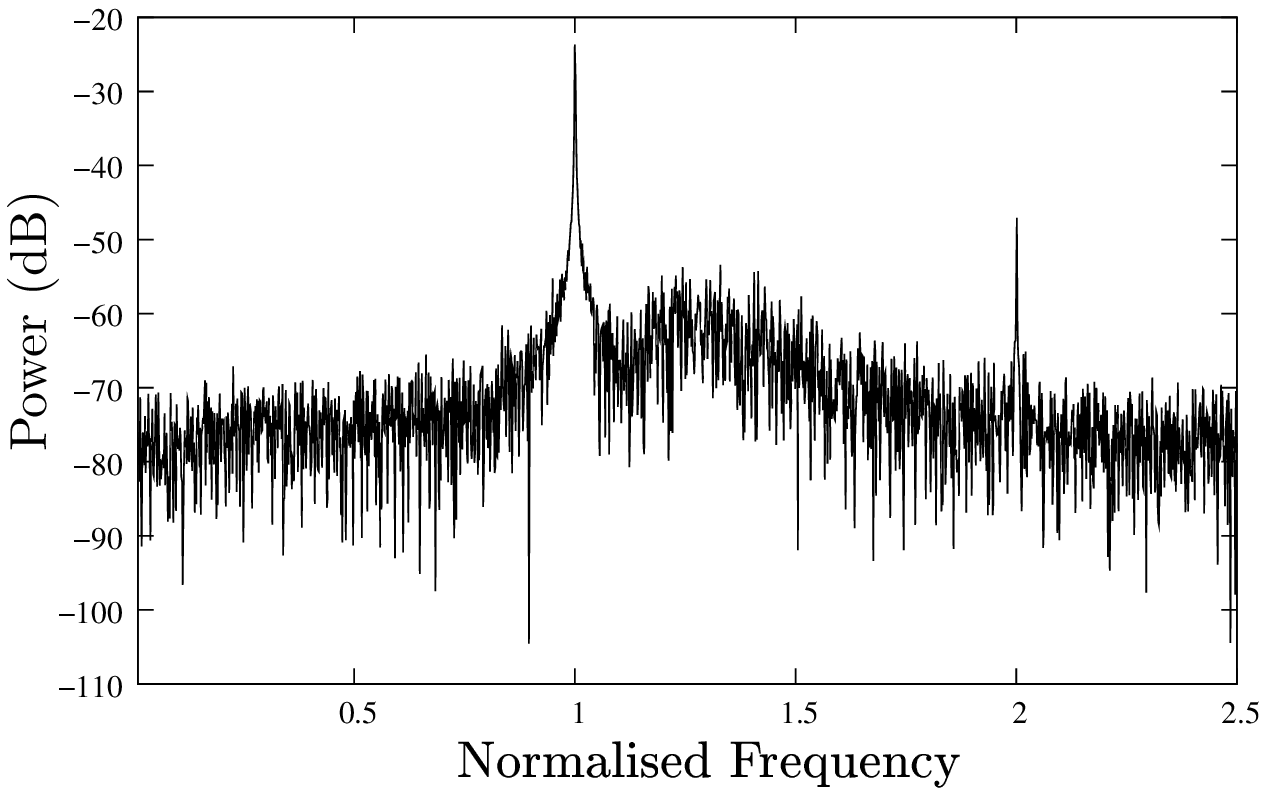}}
 }
 \subfigure[Power spectrum of $\mathcal{N}(t)$, $g=0.1$.\label{fig:ps0.1j}]{
   \resizebox*{0.45\textwidth}{!}{\includegraphics{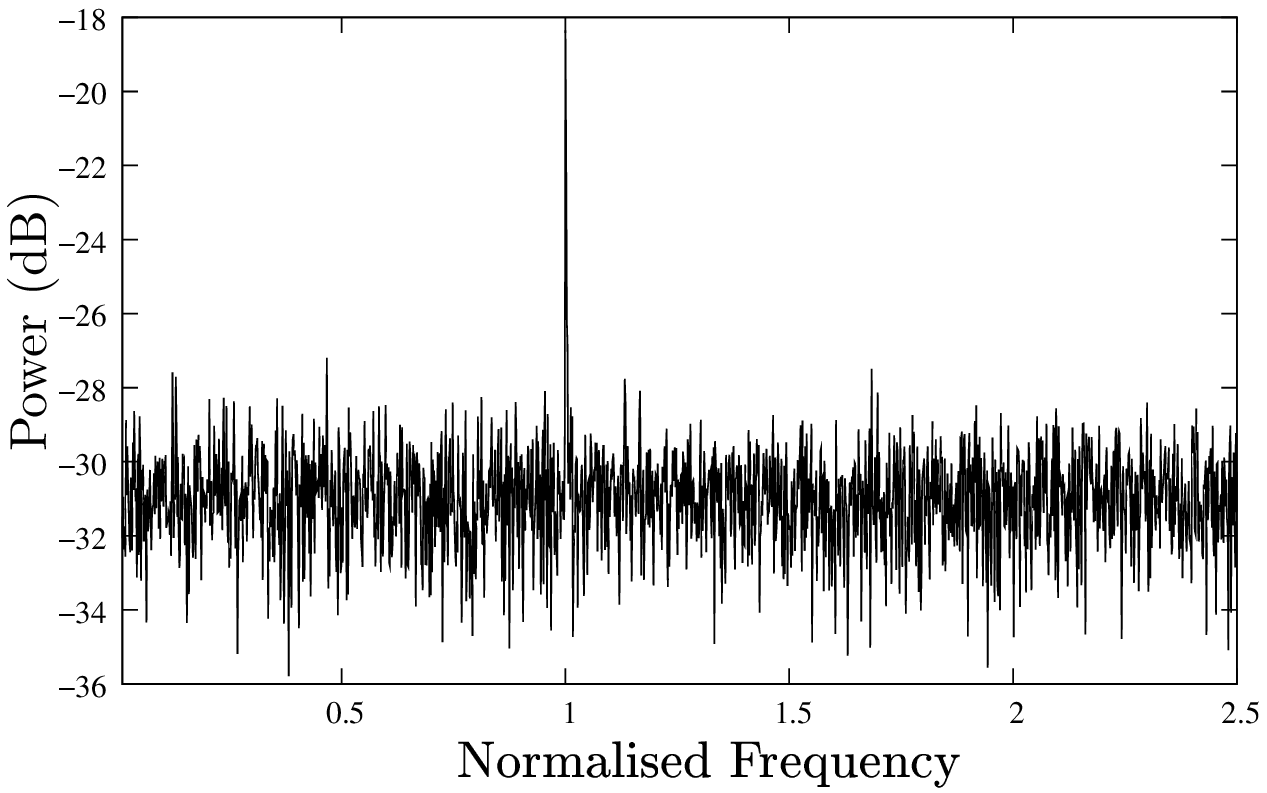}}
 }
 \subfigure[Power spectrum of $\EX{q}$, $g=0.3$.\label{fig:ps0.3x}]{
   \resizebox*{0.45\textwidth}{!}{\includegraphics{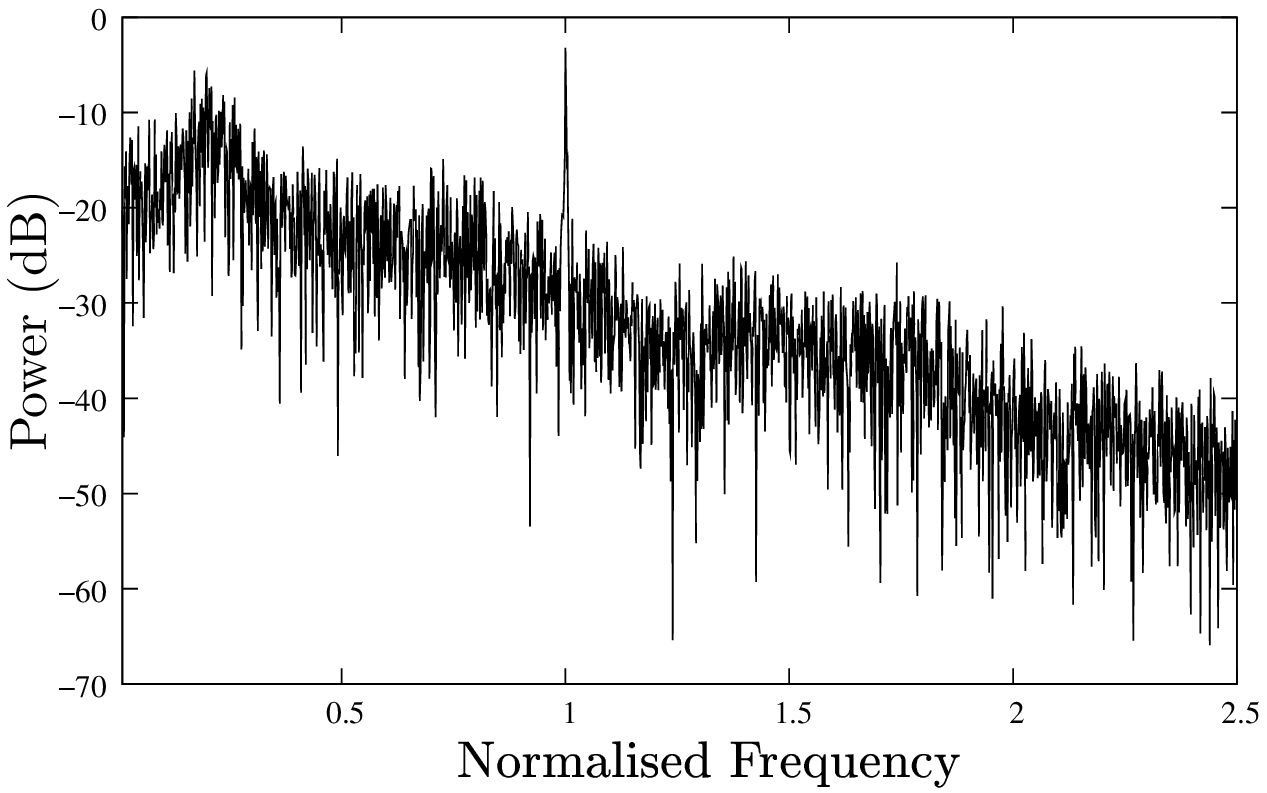}}
 }
 \subfigure[Power spectrum of $\mathcal{N}(t)$, $g=0.3$.\label{fig:ps0.3j}]{
   \resizebox*{0.45\textwidth}{!}{\includegraphics{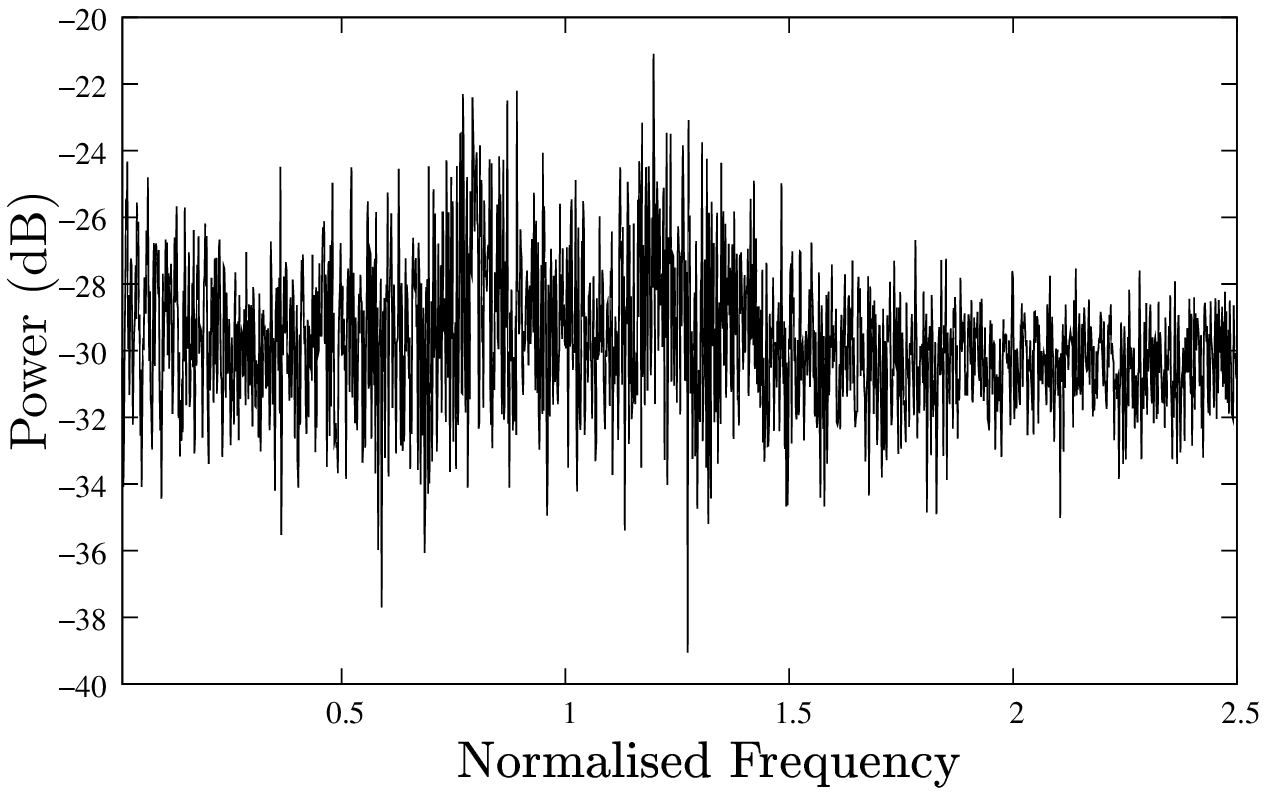}}
 }
 \subfigure[Power spectrum of $\EX{q}$, $g=1.25$.\label{fig:ps1.25x}]{
   \resizebox*{0.45\textwidth}{!}{\includegraphics{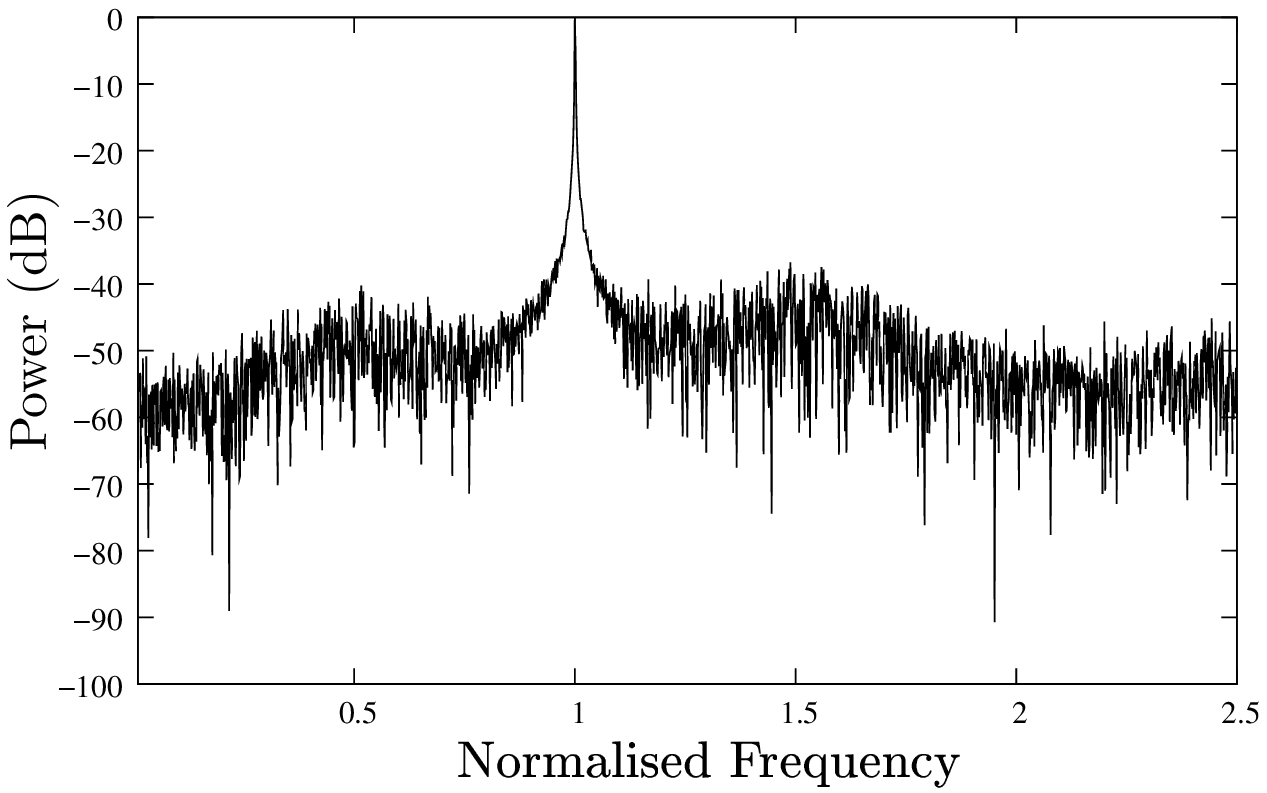}}
 }
 \subfigure[Power spectrum of $\mathcal{N}(t)$, $g=1.25$.\label{fig:ps1.25j}]{
   \resizebox*{0.45\textwidth}{!}{\includegraphics{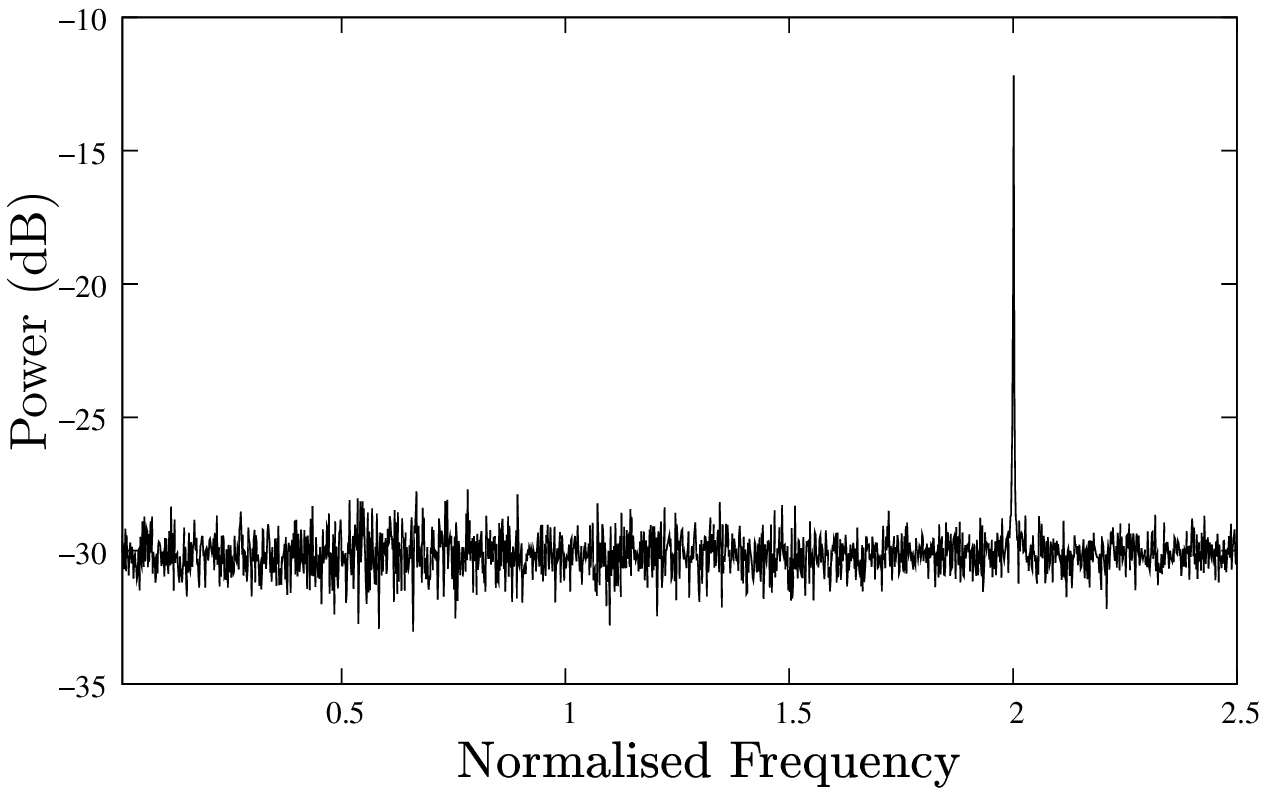}}
 }
 \subfigure[Power spectrum of $\EX{q}$, $g=2.5$.\label{fig:ps2.5x}]{
   \resizebox*{0.45\textwidth}{!}{\includegraphics{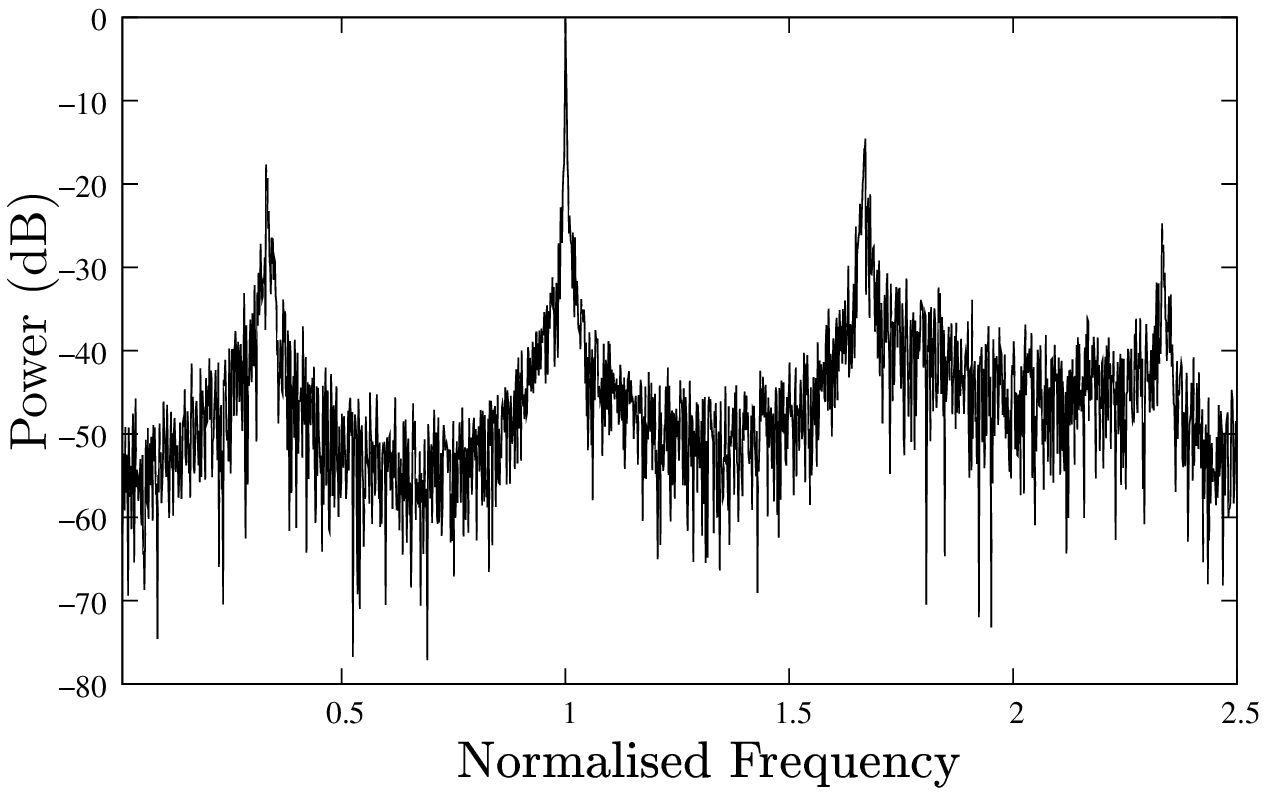}}
 }
 \subfigure[Power spectrum of $\mathcal{N}(t)$, $g=2.5$.\label{fig:ps2.5j}]{
   \resizebox*{0.45\textwidth}{!}{\includegraphics{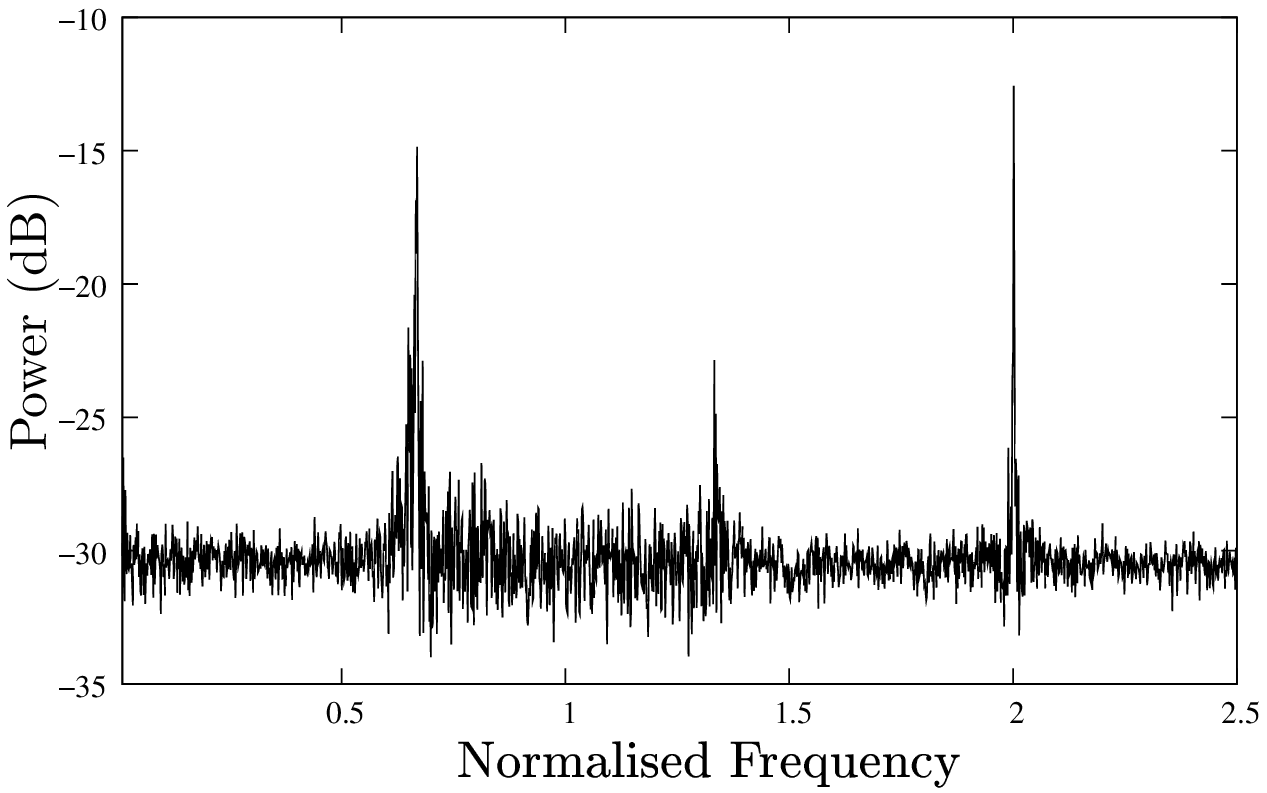}}
 }
 \caption{
   Example power  spectra for  four different
   drive amplitudes corresponding to the  regions I to IV as marked in
   the power spectrum of Fig.~\ref{fig:power3d}.
 \label{fig:ps}}
 \end{center}
\end{figure*}

 \begin{figure*}[!t]
 \begin{center}
 \subfigure[Region I - the drive $g=0.1$ (periodic stable orbit).\label{fig:phasePortraitsI}]{
   \resizebox*{0.4\textwidth}{!}{\includegraphics{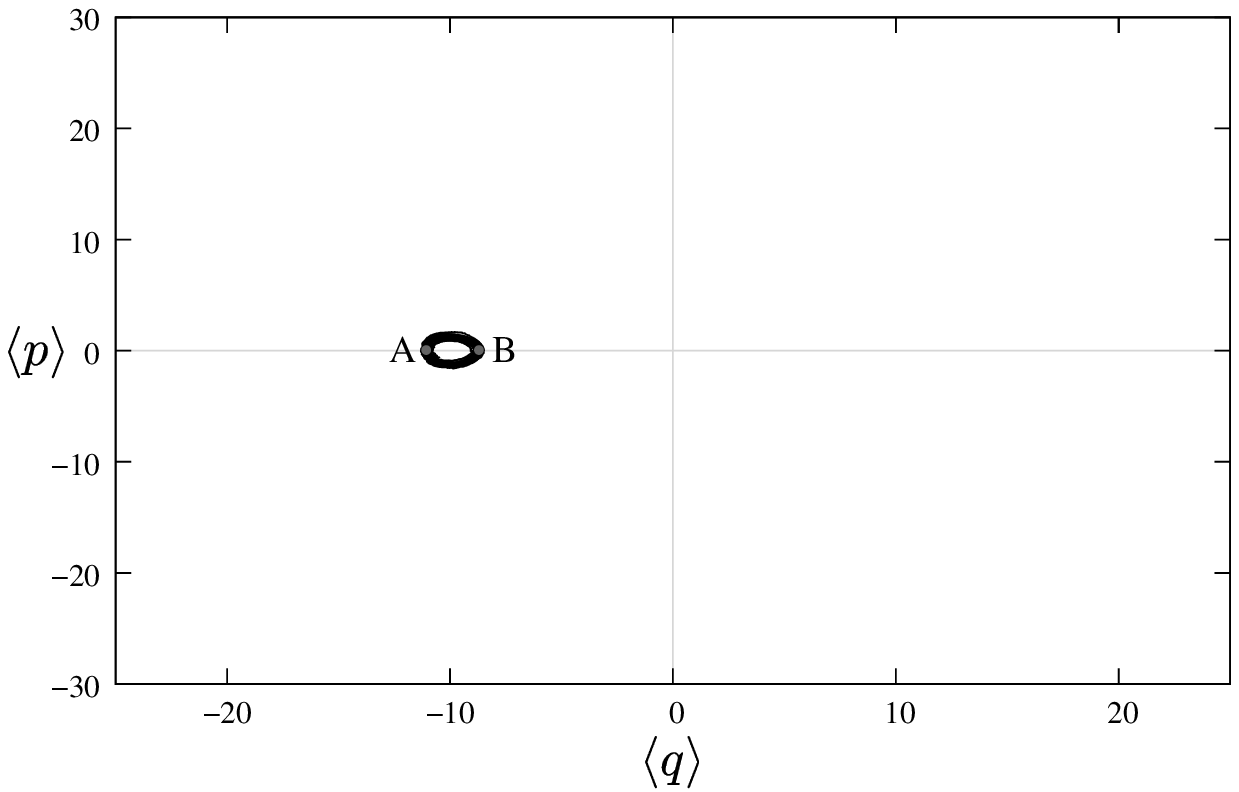}}
 }
 \subfigure[Region II - the drive $g=0.3$ (chaotic-like trajectory). \label{fig:phasePortraitsII}]{
   \resizebox*{0.4\textwidth}{!}{\includegraphics{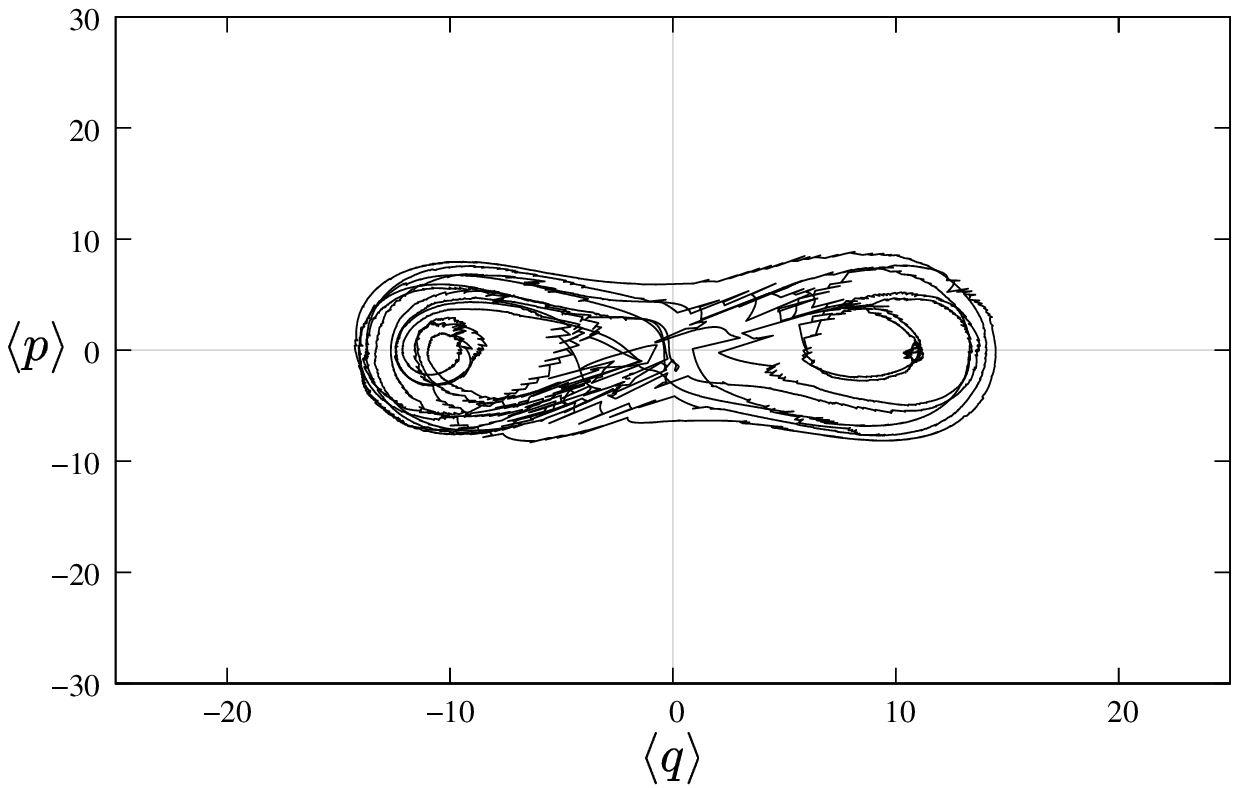}}
 }
 \subfigure[Region III - the drive $g=1.25$ (periodic stable orbit).\label{fig:phasePortraitsIII}]{
   \resizebox*{0.4\textwidth}{!}{\includegraphics{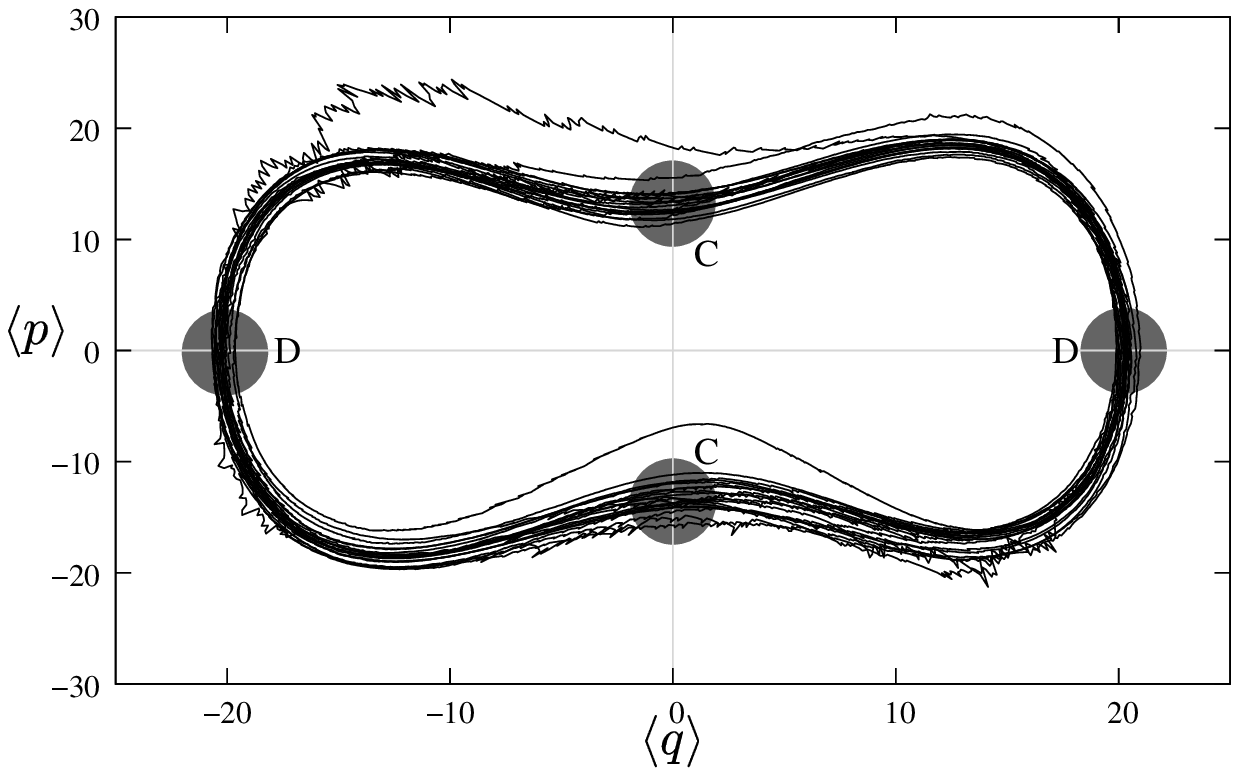}}
 }
 \subfigure[Region IV - the drive $g=2.5$ (quasi-periodic orbit).\label{fig:phasePortraitsIV}]{
   \resizebox*{0.4\textwidth}{!}{\includegraphics{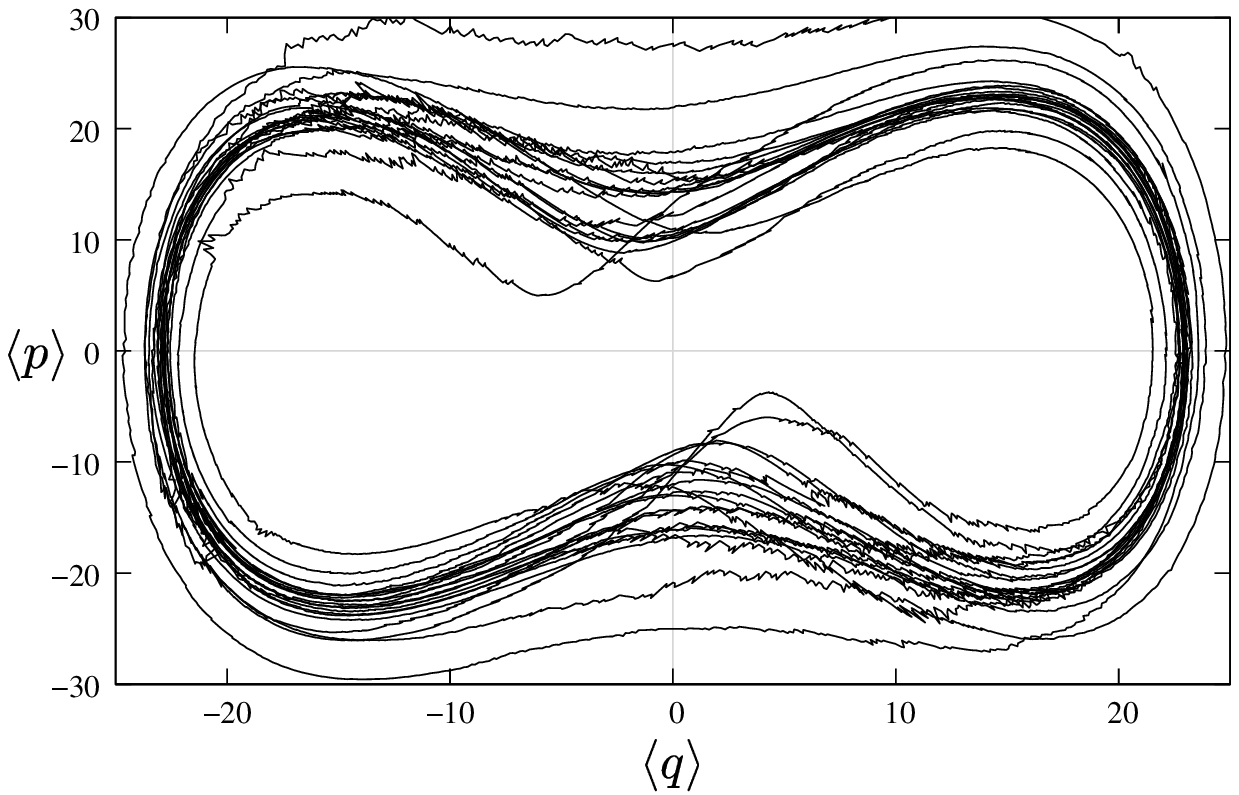}}
 }
 \caption{
   Example  phase  portraits   for  four  different  drive  amplitudes
   corresponding  to  the regions  I  to IV  as  marked  in the  power
   spectrum of Fig.~\ref{fig:power3d}.
 \label{fig:phasePortraits}}
 \end{center}
\end{figure*}
To help clarify Fig.~\ref{fig:power3d} we provide in Fig.~\ref{fig:ps}
explicit power spectra of both $\EX{q}$ and the quantum jumps
$\mathcal{N}(t)$ for regions I-IV.  As expected for region I in Figs.~\ref{fig:ps0.1x}
and~\ref{fig:ps0.1j} we see a strong resonance at the frequency of the
drive. The broadband behaviour characteristic of the chaotic phenomena
associated with region II is evident in Fig.~\ref{fig:ps0.3x} and a
concomitant, although different, structure in the power spectrum of
the detected photons.  In region III of Fig.~\ref{fig:power3d} we
again return to a periodic orbit.  In Fig.~\ref{fig:ps1.25x}, the
power spectra for $\EX{q}$ exhibits a peak at the drive frequency,
however, the power spectra of $\mathcal{N}(t)$ peaks at twice this
frequency.  The lack of coincidence between these two figures will be
explained fully in the following text.  Finally in
Figs.~\ref{fig:ps2.5x} and~\ref{fig:ps2.5j} we see power the power
spectrum of the quasi periodic dynamics of region IV, again the
discrepancy between these two figures is discussed below.

The  mechanism  through  which  the  detection of  photons  can  yield
significant information  about the  underlying dynamics of  the system
can be  understood by looking at  the phase portraits  of $\EX{q}$ and
$\EX{p}$ associated  with the regions  I--IV of Fig.~\ref{fig:power3d}
for those values  of drive used in Fig.~\ref{fig:ps},  these are shown
in Fig.~\ref{fig:phasePortraits}.
 
For  region~I there  is a  strictly  periodic response  on both  power
spectra at  the drive  frequency of the  oscillator. It can be seen from
Fig.~\ref{fig:phasePortraitsI}  that, because  of the  distance from
the origin, the  chance of there being a photon counted  at point A is
more likely than at point B.   As this occurs at the same frequency as
the  oscillations of  $\left\langle q\right\rangle  $, we  have direct
agreement in  the position of the  resonance in each  of the different
spectra.
 
In region~II, and  as is clear from Fig.~\ref{fig:phasePortraitsII},
the system is following  a chaotic-like trajectory. Although the power
spectra differ  drastically in their  structure, they do  both exhibit
broad band behaviour that is characteristic of chaotic orbits.

As  the drive  amplitude is  increased further,  region~III  in Fig.~\ref{fig:power3d} is  accessed as the behaviour observed  in region II
ceases. For  this range  of drive amplitudes  the solution is  again a
stable      periodic       orbit      as      displayed      in Fig.~\ref{fig:phasePortraitsIII}.  However,  this  time, whilst  the  power
spectrum of $\left\langle q\right\rangle $ exhibits a resonance at the
drive  frequency,  that of  $\mathcal{N}(t)$  appears  at double  this
frequency. The explanation for this  is simply that the probability of
detecting a photon  when the orbit is in a region  of phase space near
the      origin,      such     as      those      marked     C      in
Fig.~\ref{fig:phasePortraitsIII}, is less that at those further away
as in  the region of D.  This variation in probability  occurs twice a
period  and  therefore  produces  a  resonance  at  double  the  drive
frequency. An immediate corollary is that, by detecting a resonance at
either  of  these  different  frequencies  in  the  power  spectra  of
$\mathcal{N}(t)$,  we  can  determine  whether the  oscillator  is  in
region~I  or~III of  Fig.~\ref{fig:power3d}.  From our  analysis
in~\cite{everitt05} it may, in  some circumstances, be advantageous to
place the system in a chaotic  orbit. It is possible that this sort of
analysis could be used to increase or decrease drive amplitude as part
of a feedback and control element for quantum machinery.

Finally,  the  power spectrum  of  $\left\langle  q\right\rangle $  in
region~IV   of Fig.~\ref{fig:power3d}~(a)   is   characteristic  of
quasi-periodic behaviour.  Using a similar argument to  the one above,
we can transfer these features  onto the spectrum of $\mathcal{N}(t)$. 
If  we compare  this result  with  the, albeit  noisy, phase  portrait
of Fig.~\ref{fig:phasePortraitsIV}  there   is  clear  evidence  of
quasi-periodic behaviour.

   We have  demonstrated, using the Duffing oscillator  as our example
     system,  that  the  different  features exhibited  in  the  power
     spectrum of  the photon count can be  associated with concomitant
     features in the power spectrum of the position operator (and vice
     versa).   We note that  for any  given experimental  system where
     there is  a direct correspondence  between the power  spectrum of
     $\mathcal{N}$ and \EX{x} that the power spectrum of $\mathcal{N}$
     provides  us  with  the  same  amount of  information  about  the
     underlying  dynamics (e.g. chaotic,  quasi-periodic etc.)  as the
     power spectrum  of $\EX{x}$. We  would like to emphasise  that if
     this  direct  correspondence did  not  exist  then  we would  not
     necessarily be able to make  such an assertion. For example, this
     situation  might occur for  a system  in which  there was  a high
     degree    of    symmetry    in    the    $\EX{x}-\EX{p}$    phase
     portrait. However, such  a detailed study is beyond  the scope of
     this paper.

\section{Conclusion}

In this work we have shown that, via analysis of the power spectra of
the photons detected in a quantum jumps model of a Duffing oscillator,
we can obtain signatures of the underlying dynamics of the oscillator.
Again, we note that the decoherence associated with actually measuring
these jumps is that which, through localisation of the state vector,
enables these classical-like orbits to become manifest.  We have also
demonstrated that the power spectra of the counted photons can be used
to distinguish between different modes of operation of the oscillator.
Hence, this or some form of time-frequency analysis, could be used in
the feedback and control of open quantum systems, a topic likely to be
of interest in some of the emerging quantum technologies.

\begin{acknowledgments}
  The  authors  would like  to  thank  T.P.~Spiller  and W.~Munro  for
  interesting  and informative  discussions.  MJE would  also like  to
  thank P.M.~Birch for his helpful advice.
\end{acknowledgments}

\bibliography{references}

\end{document}